# Clinical Trial Drug Safety Assessment for Studies and Submissions Impacted by COVID-19

Mary Nilsson[a], Brenda Crowe[a], Greg Anglin[b], Greg Ball[c], Melvin Munsaka[d], Seta Shahin[e], Wei Wang[b]

[a]Eli Lilly and Company, Indianapolis, IN USA; [b]Eli Lilly Canada Inc., Toronto, Ontario, Canada; [c]Merck & Co., Inc., Rahway, NJ, USA; [d]AbbVie Inc., North Chicago, IL, USA; [e]Amgen Inc., Thousand Oaks, CA, USA

*Correspondence:*

Mary Nilsson

Lilly Corporate Center

Indianapolis, IN  46285

317-651-8041

nilsson_mary_e@lilly.com

# Clinical Trial Drug Safety Assessment for Studies and Submissions Impacted by COVID-19

ABSTRACT

In this paper, we provide guidance on how standard safety analyses and reporting of clinical trial safety data may need to be modified, given the potential impact of the COVID-19 pandemic. The impact could include missed visits, alternative methods for assessments (such as virtual visits), alternative locations for assessments (such as local labs), and study drug interruptions. We focus on safety planning for Phase 2-4 clinical trials and integrated summaries for submissions. Starting from the recommended safety analyses proposed in white papers and a workshop, created as part of an FDA/PHUSE collaboration (PHUSE 2013, 2015, 2017, 2019), we assess what modifications might be needed.

Impact from COVID-19 will likely affect treatment arms equally, so analyses of adverse events from controlled data can, to a large extent, remain unchanged. However, interpretation of summaries from uncontrolled data (summaries that include open-label extension data) will require even more caution than usual. Special consideration will be needed for safety topics of interest, especially events expected to have a higher incidence due to a COVID-19 infection or due to quarantine or travel restrictions (e.g., depression). Analyses of laboratory measurements may need to be modified to account for the combination of measurements from local and central laboratories.

# 1 INTRODUCTION

The pandemic of coronavirus disease (COVID-19) has had broad impact on ongoing clinical trials. Guidance has been released by various organizations and regulatory agencies, e.g., Association of Clinical Research Organizations (2020), European Medicines Agency (2020a, 2020b), McDermott and Newman (2020), U.S. Food and Drug Administration (2020) to address some of the challenges. As indicated by these guidance documents, challenges may arise from quarantines, site closures, travel limitations, or other considerations if site personnel or trial subjects become infected with COVID-19 conditions. These challenges may lead to difficulties in meeting protocol-specified procedures, including administering or using the investigational product or adhering to protocol-mandated visits and laboratory/diagnostic testing. Thus, study drug interruptions could be more common and longer in duration, and missed visits and patient discontinuations could be more common. Alternative methods for safety assessment could be implemented, e.g., phone contact, virtual visits, alternative locations for assessment (including local labs or imaging centres), that lead to differences in how patient information is received and recorded. The method of obtaining information should be considered carefully as there may be limitations in interpretation depending on the collection approach (PHUSE 2017).

PHUSE (2013, 2015, 2017, 2018, 2019), in a collaborative effort with the FDA (Rosario et al. 2012), has provided recommendations for standard safety analyses for clinical study reports and integrated submissions, contained in a collection of white papers and a workshop. One of the primary purposes of the standard safety analyses is to identify adverse events or changes in laboratory measurements, vital signs, or ECGs that require further scrutiny for adverse drug reaction (ADR) determination. ADRs are undesirable effects reasonably likely to be caused by a study drug. The standard safety analyses provide key information when determining ADRs for the investigational product, but are not the only factors (CIOMS Working Groups III and V 1999, CIOMS Working Group VI 2005, PHUSE 2019).

Using these standard safety analyses as a framework, in this paper, we examine the potential impact of COVID-19 on the scientific evaluation of safety data from clinical trials overlapping in time and geography with the pandemic. Guidance is provided on how to simply and properly reframe the analyses. We have chosen the recommendations from PHUSE white papers and workshop as our starting point, as they likely reflect the types of analyses that are planned for many ongoing studies. Whether or not product teams have implemented plans in accordance with the PHUSE recommendations, the principles should still apply.

The scope of this paper is most applicable for Phase 2-4 ongoing clinical trials in indications unrelated to infections and respiratory diseases. Specific issues related to anti-infectives and respiratory drugs, and studies that are launching now to treat the COVID-19 infection itself are out-of-scope.

We understand that these extraordinary circumstances might provoke additional interesting questions. For example, are there differences in event reporting by patients in virtual visits versus live visits? However, for clinical study reports and submissions, we recommend against reporting of analyses that would distract from the main focus of establishing the safety profile of investigational product.

We recognize that when assessing the potential impact of COVID-19 on standard safety analyses, discussion could evolve to consider updating safety planning to use the estimand framework (Unkel et al. 2019) if not already incorporated, and/or to use alternative methods proposed in recent literature e.g., Unkel et al. (2019), Stegherr et al. (2019) and references therein. Unless modifications are needed to address insufficiencies in the current statistical analysis plans, we recommend against making major changes.

This paper has been prepared by statisticians to meet the emergent need to provide guidance on how to reframe the most common analyses of clinical safety data that may have been impacted by the COVID-19 pandemic. A follow-on paper may be prepared to provide additional details and considerations.

## 2  DATA COLLECTION ASSUMPTIONS

This paper is not intended to provide details on modifications to data collection that might be needed in ongoing studies impacted by COVID-19. However, we do make the following assumptions:

- There will be a way to identify patients who have visits impacted by COVID-19 and the way in which patients are identified will be included in the Statistical Analysis Plan for the study.
- If patients cannot attend visits due to infection, quarantine, or travel restrictions, key safety data collection (adverse events [AEs], serious adverse events [SAEs], critical labs) will continue through alternative means (such as through phone contacts, virtual visits, and local labs). In the case that a longer-than-usual time elapsed since safety data were collected, patients may remember fewer AEs, but should remember the most impactful events.

For recommendations on how to represent data changes in studies impacted by COVID-19, consult the Clinical Data Interchange Standards Consortium (CDISC) COVID-19 interim guide (https://wiki.cdisc.org/display/COVID19/CDISC+Interim+User+Guide+for+COVID-19).

## 3  ASSESSING COVID-19 IMPACT

As noted earlier, multiple sources have examined the potential impacts of COVID-19 upon clinical trial data.  Generally speaking, the biggest impacts are due to quarantines and stay-in-place orders leading to additional discontinuations, missing data due to missed visits, treatment interruptions, or procedures performed differently to enable remote assessments.

In evaluating safety data, a fundamental concern is whether there are any variables that might apply differentially across treatment groups, in a way that could influence the conclusions.  Such variables include patient characteristics (such as sex, age, race) but also include aspects of study conduct (such as discontinuations, missed visits, protocol deviations, number of missed doses).  For example, discontinuations should be examined, and not only with regard to proportions of discontinuations in each treatment group, but also with respect to any patterns in timing of discontinuations and followup time.  The potential impact of any differences should be considered.

In the context of characterizing impacts from COVID-19, evaluation of these patient characteristics and other aspects still are relevant and appropriate.  However, additional evaluations may be warranted, such as assessing the proportion of patients in some way impacted by COVID-19 and the proportion of visits performed remotely rather than in person, in order to decide if there are meaningful differences across treatment groups.  The general expectation is that, while impacted patients may differ from patients not impacted, that there would not be any considerable differences among treatment groups in these characteristics.

To summarize, additional analyses may need to be performed in order to make decisions about the adequacy of current plans.  In many cases, COVID-19 will have impacted each treatment arm similarly and the originally planned safety analyses may proceed without modification, at least for the purposes of identifying ADRs.  For estimation of incidence of events, additional methods may be needed, depending on the extent of the missing data. (See Sections 5.2 and 9 for further information).

# 4 GENERAL SUMMARY OF CONCOMITANT MEDICATIONS

As noted in PHUSE 2018, it is recommended to summarize concomitant medication use between treatment arms. The primary purpose of this summary is to assess whether there is an imbalance between concomitant medications among treatments that would be important to consider when reviewing adverse event summaries. With the COVID-19 pandemic, it's quite likely that there would be changes in the concomitant medication use. However, since the focus on this summary is to detect imbalances, the planned summary and associated purpose does not need to change. A separate summary of medications used to treat COVID-19 would generally not be warranted.

# 5 GENERAL ANALYSES OF ADVERSE EVENTS

## 5.1 *Comparing percentages between treatments*

As noted in PHUSE (2017), for fixed-duration studies with similar distribution of follow-up times among treatment groups, comparing percentages of patients with specific treatment-emergent adverse events (TEAEs), SAEs, and AEs leading to study drug discontinuation among treatment arms (such as for investigational product versus placebo) is generally useful and commonly planned for helping to decide whether an event is an ADR. The intent of these analyses is to assess the imbalance between treatment arms and the magnitude of effect, which are 2 of several important factors to consider when deciding if an event is an ADR (Crowe et al. 2013, Ma et al. 2015, PHUSE 2019). If substantially more study patients discontinue study or study treatment during the controlled period in one treatment arm versus another, then different analytical approaches are needed. See Section 10.9 of PHUSE (2017) and Stegherr et al. (2019) (and the references therein) for a discussion of pitfalls of percentage-based methods and possible alternative methods. With the COVID-19 pandemic, it can be expected that more patients will discontinue early or have periods in which study drug has been interrupted. Unless the impact is considerably different across treatment arms, analytical plans can generally remain unchanged.

## 5.2  *Summarizing Event Data without a Control*

Within statistical analysis plans for safety, there are often plans to summarize counts and percentages and/or exposure-adjusted incidence rates (EAIRs) for adverse events for the investigational product, without a control arm. This is common for studies with an extension period or extension studies. Generally, the intent of these summaries is to provide an easy way to identify some of the rarer events that might require case review. These summaries are not usually used for any comparisons. If they are used for comparisons, caution is required. This would be true even before introducing any issues arising due to the COVID-19 pandemic. With the COVID-19 pandemic, the issues associated with comparing uncontrolled data to other sources might be exaggerated. For example, with the COVID-19 pandemic, there could be a substantial amount of time in which patients are off investigational product (e.g., during a quarantine). Percentages and EAIRs could therefore be under-estimated. As another example, percentages and EAIRs may be impacted due to differences in ascertainment of adverse events (e.g., phone call instead of a site visit). Additionally, events associated with COVID-19 (e.g., fever, cough) or events associated with physical or social isolation (e.g., depression) might appear at a more frequent rate. Consequently, comparing an EAIR from studies/integrated summaries impacted by COVID-19 with an EAIR from the literature or other source could be even more problematic than usual. When it is necessary to compare an EAIR with another source to use as a background rate, summarizing up to

COVID-19 impact or by COVID-19 subgroups (such as patients without impact and patients with impact) might be helpful for ADR decision-making.

# 6  GENERAL ANALYSES OF LABORATORY DATA

When a central lab is normally used for a study and local labs are subsequently used due to COVID-19, this is an important impact of COVID-19 to consider.  This situation can occur if patients are unable to attend a site visit (for example, the site is at a hospital that is closed to clinical trial activities) but are able to go to a different location for select laboratory measurements needed for safety monitoring.

If a local lab is sometimes used but the measurements are not brought into the study database, then analyses using central lab data will be conducted with less complete data than what would have otherwise been available.  (As noted in Section 2, we assume that we will have laboratory data for critical labs).  Since analyses using central lab data will be incomplete, interpretation may need to rely on a combination of analyses based on lab measurements and adverse event summaries to a greater degree than usual.

If a local lab is used and the measurements are brought into the study database, then analytical plans for central tendency analyses will need to be updated to provide clarity on if and when local lab measurements and central lab measurements will be combined. For many lab analytes, combining local and central labs could help fill in the gaps, providing more complete data.  However, for some analytes, directly combining the data may not be appropriate. Note, even when combining local and central labs, additional variability and uncertainty can be added into the data.  See Section 6.2 for more details.

### 6.1 Comparing percentages of shifts to high/low between treatments

As noted in PHUSE (2015), comparing percentages of patients shifting from normal/low to high and normal/high to low (sometimes referred to as treatment-emergent highs and lows) among treatment arms is recommended. Similar to comparing percentages for adverse event data, the intent of these analyses is to assess the imbalance among treatment arms. As discussed in Section 3, with the COVID-19 pandemic, perhaps more patients will discontinue early or have periods in which study drug has been interrupted, however, the impact will likely be similar across treatment arms. Thus, analytical plans can mostly remain unchanged. For these summaries, combining measurements from local labs and central labs is generally appropriate, as long as the limits from the associated lab are used.

### 6.2 Boxplots by visit with simple summary statistics

As noted in PHUSE (2013), summarizing changes over time by treatment using simple statistics via boxplots is recommended for individual studies. Using simple summary statistics could be problematic if data collection is impacted by COVID-19. During the pandemic there could be a substantial number of missed visits. Under these circumstances, reporting means based on a mixed model for repeated measures (MMRM) instead of simple means may be more appropriate. Moreover, if a local laboratory is used and the measurements are brought into the study database, the study team will need to decide which laboratory measurements can be combined. Alternatively, study teams can choose a different analytical approach that allows for combining laboratory measurements from different laboratories. See Section 6.2.4 from PHUSE (2013).

### 6.3 *Comparing changes to minimum/maximum values between treatments*

As noted in PHUSE (2013), comparing change to a minimum/maximum value between treatments is generally recommended for integrated summaries. As with comparing percentages between treatments, comparing changes to minimum/maximum values between treatments should be appropriate, unless the average number of measurements is very different among treatment arms. If local labs are used and the measurements are brought into the study database, the same considerations described in Section 6.2 apply for these analyses.

### 6.4 *Hepatotoxicity*

Typically, in submissions, there is an expectation to assess the potential for drug-induced hepatic injury (FDA 2009). As part of this evaluation, a plot of alanine aminotransferase (ALT) versus total bilirubin is often created (Senior 2014). The upper right quadrant (>3X ULN ALT, >2X ULN total bilirubin) is usually referred to as "Hy's Law Range" or "Potential Hy's law cases". Identification of true Hy's Law cases require additional considerations, but this plot can be used to graphically show whether a study drug has the potential to cause hepatic injury. For this assessment, every occasion of having an ALT >3X ULN and total bilirubin >2X ULN matters. If local labs are used and data are not brought into the study database, there's a potential for missing patients that would otherwise have been in the Hy's Law Quadrant. Careful review of adverse event data would be required. Certainly, it would be better if all the results for hepatic enzymes are brought into the study database. If local labs are used and data are brought into the study database, the limits from the local laboratory should be used to determine the multiple of the upper limit of normal.

# 7   INTRINSIC FACTORS

For large individual studies and integrated summaries of safety, there are often plans to summarize percentages of common TEAEs by subgroups.  These subgroups usually include gender, age categories, and race.  Additional subgroups may be added depending on the indication under study.  See Figure 12.2 of PHUSE (2017).  Summarizing by COVID-19 subgroups for common TEAEs or other general safety outcomes will generally not be informative for ADR decision-making and are unnecessary.

# 8   CASE REVIEWS

In addition to assessing imbalances with a control and magnitude of effect, ADR determination includes several other factors, including impressiveness of individual cases.  Case reviews are conducted to assess the potential relationship with investigational product versus a concomitant medication versus other conditions the patient may be experiencing.  It's common to create individual patient displays (such as narratives and graphical patient profiles) to facilitate this review.  These displays usually include demographics, study drug exposure, concomitant medications, AEs, labs, vital signs, and – when applicable – ECGs.  For exposure, it's common to show start and stop dates of study drug.  If study drug has been interrupted due to a COVID-19 quarantine or other reason, this should be reflected.  If a visit has been impacted by COVID-19 in any manner, this should be reflected.  Knowing dates for study drug exposure and knowing whether visits have been impacted in any manner would be helpful for these case reviews.

# 9   SAFETY TOPICS OF INTEREST

While analytical planning for general safety assessment can largely remain the same

(with some exceptions), special consideration is needed for safety topics of interest, particularly those that could have a higher incidence due to COVID-19 infection or due to the physical or social isolation caused by mandates to stay at home (depression, for example). The cross-disciplinary team should discuss the possibility for additional or alternative methods that might be warranted. For example, summaries up to COVID-19 impact or by COVID-19 subgroups for some safety topics of interest are likely warranted. Additionally, more complex methodology (such as Kaplan Meier plots, Cox proportional hazards methods, and/or competing risk models) may need to be implemented. The need for additional methods will depend on the safety topic of interest and the extent of the COVID-19 impact (and impact from other factors). If choosing across alternative methods, it is important to try to connect the method with the eventual interpretation, and to understand the pros and cons of the methodological choices. A full discourse on these methods is beyond the scope of this paper. Nevertheless, we offer some insights on one particular methodology, namely, competing risks. COVID-19 logistical problems are unlikely to introduce a need for competing risk analysis for a study for which no competing risk analysis was needed prior to the COVID-19 pandemic, unless a study has many deaths from COVID-19 infection. Competing risks are events that preclude or greatly alter the occurrence of the main event of interest. For example, if the event of interest is myocardial infarction, death from other causes would be considered a competing risk. Competing risks are different from other concurrent events in that they actually preclude the event of interest from happening, whereas events like early study discontinuation prevent the event from being *observed*. Various authors, e.g., Allignol et al. (2016), Bender et al. (2016), Geskus (2016), Hengelbrock et al. (2016), Proctor and Schumacher (2016), Unkel et al. (2019) have written about the need to consider competing risks in the assessment of the risk/probability of adverse events. As noted by Allison (2018), standard Kaplan Meier or Cox proportional hazards methods perform better for determining whether or not the drug is causally related to the AE than methods that take into account the competing risk. Alternatively, if interest is in getting an accurate percentage of patients with the event (as for the product label), estimation methods that take competing risks into account may be useful.

For some compounds, such as those used to treat auto-immune conditions, infections might already be a safety topic of interest. For these compounds and possibly others, considering the COVID-19 infection itself as a safety topic of interest could be warranted. This could arise if there is biological plausibility for a greater risk for infection and/or if there is imbalance in COVID-19 infections between treatment arms. Additional analyses that might be needed in this situation are out-of-scope, but would likely follow similar approaches to any safety topic of interest.

## 10 PRODUCT LABELING

For product labeling purposes, additional displays are often needed to characterize ADRs. These often include an assessment of event duration or whether a change in labs/vitals is transient versus persistent. With the COVID-19 pandemic, it's possible that less information will be available to assess these patterns in the data due to missing visits, early study discontinuation, early study drug discontinuation, or study drug interruption. Using group means to assess transient (versus persistent) patterns always has the potential to be misleading (PHUSE 2019), but the potential is even greater if discontinuations or interruptions are more common. Thus, for these assessments, using a display that graphically displays individual patient data is recommended. For events, see Appendix B in (PHUSE 2017) for an example of a plot showing events over time

(onset and duration). For labs/vitals, a spaghetti plot (plot of values [vertical axis] versus time [horizontal axis] and connecting the dots chronologically with lines for each patient) serves this purpose. In a spaghetti plot, symbols or color can be used for when a patient is on or off drug.

When communicating about ADRs in labeling, cautionary language on the limitations of comparing with other labels is usually included. For compounds in which there is a large impact from COVID-19, the cautionary language may need to be expanded to mention the potential for under- or over-reporting due to COVID-19. Furthermore, depending on the rarity of the event and the extent of COVID-19 impact on the study, it's possible that it would be more appropriate to use a percentage from the non-COVID-impacted group. It may be useful to have these available.

## 11 SUMMARY OF SAFETY EVALUATION

Table 1 provides a summary of the recommendations that are applicable for most situations. However, the details in previous sections are needed to fully understand the recommendations, possible exceptions, and cautionary notes. The analysis type included in this table are from select analyses described in the PHUSE white papers.

*Table 1. Summary of recommendations for common safety analyses (as recommended by PHUSE white papers and safety workshop (PHUSE 2013, 2015, 2017, 2018, 2019)) for clinical trials impacted by COVID-19*

| Analysis Type | Recommendation for updating analysis plan |
|---|---|
| Concomitant medications - Comparing percentages between treatments | Likely no change needed |
| TEAEs, SAEs, AEs leading to study drug discontinuation - comparing percentages between treatments | For general safety summaries, likely no change needed |
| TEAEs for uncontrolled data – Summarizing counts, percentages, exposure-adjusted incidence rates | For general safety summaries, likely no change needed |
| Labs/vitals – Comparing percentages between treatments (e.g., treatment-emergent highs and lows) | No change needed, except if data from local labs are included in the clinical trial database, low/normal/high should be determined using the local lab reference limits |
| Labs/vitals - Boxplots by visit with simple summary statistics under the boxplot (recommended for individual studies) | If COVID-19 impact includes a lot of missed visits, consider reporting means based on MMRM instead of the simple mean under the boxplot. If data from local labs are included in the clinical trial database, a decision is needed on which lab analytes can use combined data versus not or change to an alternative method that allows for the combination |
| Labs/vitals - Comparing mean change to minimum/maximum values between | No change needed, except if data from local labs are included in the clinical trial database, a decision is needed on which |
| treatments (recommended for integrated summaries) | lab analytes can use combined data versus not or change to an alternative method that allows for the combination |
| Hepatotoxicity – Plot of alanine aminotransferase versus total bilirubin | If local labs are used and data are not brought into the study database, there's a potential for missing patients that would otherwise have been in the Hy's Law Quadrant.  Careful review of adverse event data would be required.  If local labs are used and data are brought into the study database, use the limits from the local lab to determine the multiple of the upper limit of normal. |
| Intrinsic Factors:  Subgroup analyses for common TEAEs | Likely no change needed |
| Case reviews | Include study drug exposure start/stop dates and information on visits impacted by COVID-19 |
| Safety topics of interest | For safety topics of interest that could have a higher incidence due to COVID-19 infection or due to the physical or social isolation caused by mandates to stay at |

| | home, summarizing by COVID-19 subgroups (e.g., patients without impact, patients with impact) might be helpful for ADR decision-making. |
|---|---|
| ADR characterization in product labeling (e.g., event duration, transient versus persistent assessment) | If group summaries are planned, consider replacing with patient-based displays |
| ADR communication (e.g., percentages to report) in product labeling | If the COVID-19 impact is large, for some ADRs, reporting the percentage from the non-COVID-impacted group might be warranted. Cautionary language may need to mention the potential for under- or over-reporting due to COVID-19. |

## 12 CONCLUDING REMARKS

For general assessment of AEs, SAEs, and AEs leading to permanent discontinuation of study drug (for controlled data), we believe analysis plans can largely remain unchanged unless COVID-19 logistics introduce differential observation time between treatments. For safety topics of interest expected to have a higher incidence due to a COVID-19 infection or due to quarantine or stay-at-home mandates (such as for depression), comparing exposure-adjusted incidence rates from uncontrolled data with a background rate from literature or alternative source could be more problematic than usual. For such safety topics, limiting data up to COVID-19 impact or summarizing by COVID-19 subgroups (for example, patients who had study visits impacted by COVID-

19 versus patients who did not have any study visits impacted by COVID-19) should be considered. Conducting general safety analyses by COVID-19 subgroups does not seem warranted. For analyses of laboratory measurements, analysis plans will likely need to be updated if there is a combination of measurements from local labs and central labs in the study database. When communicating ADRs in labelling, cautionary language on the limitations of comparing with other labels may need to be expanded to mention the potential for under- or over-reporting due to COVID-19 impact.